\documentclass[12pt]{article}
\usepackage{euscript,amsfonts}
\begin{document}
\title{\Large\bf{Quantum BRST charge and $OSp(1|8)$ superalgebra of twistor-like $p$-branes with exotic supersymmetry and Weyl symmetry}}
\author{D.V.~Uvarov${}^a$ and A.A.~Zheltukhin${}^{a,b}$\\
{\normalsize ${}^a$ Kharkov Institute of Physics and Technology, 61108
Kharkov, Ukraine}\\
{\normalsize ${}^{b}$ Institute of Theoretical Physics, University of
Stockholm, Albanova,}\\
{\normalsize  SE-10691 Stockholm, Sweden}}
\date{}
\maketitle

\begin{abstract}
Algebra of the constraints of twistor-like $p$-branes restoring $\frac34$ fraction of the spontaneously broken $D=4$ $N=1$ supersymmetry is studied using the conversion method. Classical and quantum realizations of the BRST charge, unified superalgebra of the global generalized superconformal $OSp(1|8)$ and Virasoro and Weyl symmetries are constructed. It is shown that the quantum Hermitian BRST charge is nilpotent and the quantized $OSp(1|8)$ superalgebra is closed.
\end{abstract}

\section{\bf  Introduction}
 The physical interpretation of the central charges in supersymmetry algebra as topological  charges carried by branes \cite{agit} advanced understanding of the phenomena of partial spontaneous breaking of supersymmetry \cite{hlp}.
 Because branes are constituents of M-theory, spontaneously breaking supersymmetry, their global and local symmetries correlate with the symmetries of M-theory \cite{dufli}, \cite{hull}. Studying these symmetries resulted in the model independent  classical analysis of BPS states preserving $\frac14$, $\frac12$ or $\frac34$ fractions of the partially spontaneously broken $D=4$ $N=1$ supersymmetry \cite{gght}. A special interest to construction of a physical model with domain wall configurations preserving $\frac34$ fraction of the $D=4$ $N=1$ supersymmetry against spontaneous breaking was subscribed there. That configurations were  earlier studied  in superparticle dynamics \cite{balu} and algebraically realized as the brane intersections in \cite{gah}.
 Then the tensionless string/brane model preserving $\frac34$ fraction of the $D=4$ $N=1$ supersymmetry and generating static solutions for these tensionless objects was proposed in \cite{ZU}. These results have sharpen the general question:
whether quantum exotic BPS states saturated by the p-brane states protect the same high  $\frac{{\sf M}-1}{{\sf M}}$  fraction of $N=1$ global supersymmetry against spontaneous breaking  as in the classical case?

We have started studying the question in \cite{UZ04} on the example of the p-branes preserving $\frac34$ fraction of the partially spontaneously broken $D=4$ $N=1$ supersymmetry and found some obstacles for the quantization in the $\hat{\mathcal Q}\hat{\mathcal P}$-ordering previously studied in \cite{GLSSU} (see also \cite{HM}).
Here we analyze the quantization problem applying the BFV approach
\cite{BFV} and  construct quantum Hermitian BRST operator and the
generators of gauge Weyl, Virasoro and global $OSp(1|8)$
symmetries extended by the ghost contributions. We prove the
nilpotency of the quantum Hermitian BRST charge, its
(anti)commutativity with the quantum Hermitian generators of the
$OSp(1|8)$ superalgebra and the closure of this quantized superlgebra. At the same time we show that the
quantum $\hat{\mathcal Q}\hat{\mathcal P}$-ordered BRST
operator and the $OSp(1|8)$ generators are nonHermitian and differ
from the Hermitian ones by the presence of the divergent
terms. We discuss a possibility to overcome this obstacle by the choice of the special regularization prescription for the $p$-brane world-volume delta function $\delta^p(\vec\sigma-\vec\sigma')|{}_{\vec\sigma=\vec\sigma'}$ and its derivative $\partial_M\delta^p(0)$. A possibility of the exact cancellation of the divergent terms in the $\hat{\mathcal Q}\hat{\mathcal P}$-ordered quantum operators for other dimensions $D=2,3,4(mod 8)$ is also discussed.

\section{\bf Conversion of tensionless super $p$-brane constraints}

New models of tensionless string and $p$-branes evolving in the
symplectic superspace ${\cal M}^{susy}_{\sf M}$ and preserving
all but one fractions of $N=1$ supersymmetry were recently
studied in \cite{ZU}, \cite{ZUB}. For ${\sf M}=2^{[\frac{D}{2}]}$
$(D=2,3,4\ mod\ 8)$ the space ${\cal M}^{susy}_{\sf M}$
  extends the   standard  $D$-dimensional super space-time $(x_{ab},\theta_a) $, (where $a,b=1,2,...,2^{[\frac{D}{2}]})$ by the tensor
central charge (TCC) coordinates $z_{ab}$.  The coordinates $x_{ab}=x^m(\gamma_mC^{-1})_{ab}$ and $z_{ab}=iz^{mn}(\gamma_{mn}C^{-1})_{ab} +
z^{mnl}(\gamma_{mnl}C^{-1})_{ab}+...$ constitute components of the symmetric  spin-tensor $Y_{ab}$. In terms  of $Y_{ab}$ and the Majorana spinor $\theta_a$
the action  \cite{ZU},
invariant under the spontaneously broken $N=1$ supersymmetry and world-volume reparametrizations,
is given by
\begin{equation}\label{1}
S_p=\frac12\int d\tau d^p\sigma\ \rho^\mu U^aW_{\mu ab}U^b,
\end{equation}
where $W_{ab}=W_{\mu
ab}d\xi^\mu$ is the  supersymmetric Cartan differential one-form
\begin{equation}\label{2}
W_{\mu ab}=\partial_\mu
Y_{ab}-2i(\partial_\mu\theta_a\theta_b+\partial_\mu\theta_b\theta_a),
\end{equation}
 and $\partial_\mu\equiv\frac{\partial}{\partial\xi^\mu}$ with $\xi^\mu=(\tau,\sigma^M)$, $(M=1,2,...,p)$ parametrizing the $p$-brane world volume.
 The local auxiliary Majorana spinor
$U^a(\tau,\sigma^M)$ parametrizes the generalized momentum
$P^{ab}=\frac12\rho^\tau U^aU^b$ of tensionless $p$-brane and
$\rho^\mu(\tau,\sigma^M)$ is the world-volume vector density providing
the reparametrization invariance of $S_p$ similarly to the null branes \cite{
ZB}.

The  action (\ref{1}) has $({\sf M}-1)$ $\kappa-$symmetries
\begin{equation}\label{3}
\delta_\kappa\theta_a=\kappa_a,\quad\delta_\kappa
Y_{ab}=-2i(\theta_a\kappa_b+\theta_b\kappa_a),\quad\delta_\kappa
U^a=0,
\end{equation}
which protect $\frac{{\sf M}-1}{\sf M}$ fraction of the $N=1$ global supersymmetry to be spontaneously broken, because of the one real condition $U^a\kappa_a=0$ for  the transformation parameters $\kappa_a(\tau,\vec\sigma)
$\footnote{To remove a possible misunderstanding in this terminology let us remind that from the world-volume perspective the last fraction of the $N=1$ supersymmetry is also the symmetry of the action (\ref{1}), but it is spontaneously broken, because of the presence of Goldstone fermion $\tilde\eta=-2i(U^a\theta_a)$ encoding the single physical fermionic degree of freedom associated with $\theta_a$ (see \cite{ZUB} for details).}.

For the four-dimensional space-time the action (\ref{1}) takes  the  form
\begin{equation}\label{4}
S_p=\frac12\int d\tau d^p\sigma\ \rho^\mu
\left(2u^\alpha\omega_{\mu\alpha\dot\alpha}\bar
u^{\dot\alpha}+u^\alpha\omega_{\mu\alpha\beta}u^\beta+\bar
u^{\dot\alpha}\bar\omega_{\mu\dot\alpha\dot\beta}\bar u^{\dot\beta}\right).
\end{equation}
This action is invariant under the  $OSp(1|8)$ symmetry, which is global supersymmetry
of the massless fields of all spins in $D=4$ space-time extended by TCC
coordinates \cite{Fronsdal}, \cite{Vasiliev}.

 The Hamiltonian structure of the action (\ref{4}), described in  \cite{UZ03},  is characterized by 3 fermionic and $2p+7$ bosonic first-class constraints that generate its local symmetries, as well as, 1 fermionic and 8 bosonic second-class constraints taken into account by the construction of the Dirac bracket. 
We found that the D.B. algebra of the first-class constraints has the rank equal two and it gives rise to the higher powers of the ghosts in the BRST 
generator.

To  simplify transition to the quantum theory the conversion
method \cite{Fad}-\cite{BLS}, transforming  all the primary and
secondary constraints to the first class, has been applied in
\cite{UZ04}. To this end the additional canonically conjugate
pairs  $(P^\alpha_q, q^\alpha)$, $(\bar P^{\dot\alpha}_q, \bar
q^{\dot\alpha})$, $(P^{(\varphi)}_\tau, \varphi^{\tau})$ and the
self-conjugate Grassmannian variable $f$ have been introduced. As
a result, all the constraints have been converted to the effective
first-class constraints in the extended phase space.

The converted constraints for the auxiliary  fields  are  the following
\begin{eqnarray}\label{5}
\widetilde P^\alpha_u=P^\alpha_u+P^\alpha_q\approx0,\quad\bar{\widetilde
P}{}^{\dot\alpha}_u=\bar P^{\dot\alpha}_u+\bar P^{\dot\alpha}_q\approx0,
\end{eqnarray}
\begin{equation}\label{6}
\widetilde P^{(\rho)}_\tau
=P^{(\rho)}_{\tau}
+
P^{(\varphi)}_\tau\approx0,\quad
P^{(\rho)}_{M}\approx0.
\end{equation}
The converted  bosonic constraints
$\widetilde\Phi\equiv(\widetilde\Phi^{\dot\alpha\alpha},
\widetilde\Phi^{\alpha\beta},
\bar{\widetilde\Phi}{}^{\dot\alpha\dot\beta})$ originating from the
$\Phi$-constraints \cite{UZ04} are given by
\begin{equation}\label{7}
\widetilde\Phi^{\dot\alpha\alpha}=P^{\dot\alpha\alpha}-\tilde\rho^\tau\tilde
u^\alpha
\bar{\tilde u}{}^{\dot\alpha}\approx0,\
\widetilde\Phi{}^{\alpha\beta}=\pi^{\alpha\beta}+\frac12\tilde\rho^{\tau}\tilde u^\alpha\tilde u^\beta\approx0,\
\bar{\widetilde\Phi}{}^{\dot\alpha\dot\beta}=\bar\pi^{\dot\alpha\dot\beta}+\frac12\tilde\rho^{\tau}\bar{\tilde
u}{}^{\dot\alpha}
\bar{\tilde u}{}^{\dot\beta}\approx 0
\end{equation}
 and have  zero Poisson brackets (P.B.) with the constraints (\ref{5}), (\ref{6}) and among themselves.
The converted fermionic constraints
$\widetilde\Psi=(\widetilde\Psi^\alpha,
\bar{\widetilde\Psi}{}^{\dot\alpha})$ originating from the primary
$\Psi$-constraints and  generating four  $\kappa-$symmetries take  the form
\begin{equation}\label{8}\begin{array}{c}
\widetilde\Psi^\alpha=\pi^\alpha-2i\bar\theta_{\dot\alpha}P^{\dot\alpha\alpha}-4i\pi^{\alpha\beta}\theta_\beta+2(\tilde\rho^\tau)^{1/2}\tilde
u^\alpha f\approx0,\\[0.2cm]
\bar{\widetilde\Psi}{}^{\dot\alpha}=-(\widetilde\Psi^\alpha)^\ast=\bar\pi^{\dot\alpha}-2iP^{\dot\alpha\alpha}\theta_\alpha-4i\bar\pi^{\dot\alpha\dot\beta}\bar\theta_{\dot\beta}-2(\tilde\rho^\tau)^{1/2}\bar{\tilde
u}{}^{\dot\alpha} f\approx0,
\end{array}
\end{equation}
 where $f^\ast=f$ is an auxiliary Grassmannian  variable
characterized by the P.B.
\begin{equation}\label{9}
\{f(\vec\sigma),f(\vec\sigma')\}_{P.B.}=-i\delta^p(\vec\sigma-\vec\sigma').
\end{equation}
The addition of the field $f(\tau,\vec\sigma)$ restores the forth $\kappa$-symmetry and
transforms all $\widetilde\Psi$-constraints to the  first class.
The  Weyl symmetry constraint $\widetilde\Delta_W$
 in the extended  phase space  is
\begin{equation}\label{10}
\widetilde\Delta_W=(\tilde{\tilde P}{}^\alpha_u\tilde
u_\alpha+\bar{\tilde{\tilde P}}{}^{\dot\alpha}_u
\bar{\tilde u}_{\dot\alpha})-2\tilde\rho^\tau\tilde{\tilde
P}{}^{(\rho)}_\tau-2\rho^M P^{(\rho)}_M\approx0,
\end{equation}
where the  variables $(\tilde u^\alpha=u^\alpha-q^\alpha, \,
\tilde{\tilde P}{}^\alpha_u=\frac12(P^\alpha_u-P^\alpha_q))$ and
$(\tilde\rho^{\tau}=\rho^{\tau}-\varphi^{\tau}, \, \tilde{\tilde
P}{}^{(\rho)}_{\tau}=\frac12(P^{(\rho)}_\tau-P^{(\varphi)}_\tau))$ form canonically
conjugate pairs \cite{UZ04}.
Finally, the converted constraints
$\widetilde L_{M}$ of the world-volume
$\vec\sigma-$reparametrizations are
\begin{equation}\label{11}
\begin{array}{rl}
\widetilde
L_{M}=&\partial_{M}x_{\alpha\dot\alpha}P^{\dot\alpha\alpha}+\partial_{M}z_{\alpha\beta}\pi^{\alpha\beta}+\partial_{M}
\bar
z_{\dot\alpha\dot\beta}\bar\pi^{\dot\alpha\dot\beta}+\partial_{M}\theta_\alpha\pi^\alpha+\partial_{M}\bar\theta_{\dot\alpha}\bar\pi^{\dot\alpha}
\\[0.2cm]
+&\partial_M\tilde u_\alpha\tilde{\tilde P}{}^\alpha_u+\partial_M\bar{\tilde u}_{\dot\alpha}\bar{\tilde{\tilde
P}}{}^{\dot\alpha}_u-\tilde\rho^{\tau}\partial_M
\tilde{\tilde P}{}^{(\rho)}_{\tau}-\rho^{N}\partial_M
P^{(\rho)}_{N}-\frac{i}{2}f\partial_M f\approx0.
\end{array}
\end{equation}
 The P.B.  superalgebra of the converted
first-class constraints (\ref{5})-(\ref{8}), (\ref{10}), (\ref{11})  is described by the
following non zero relations
\begin{equation}\label{12}
\{\widetilde\Psi^\alpha(\vec\sigma),\widetilde\Psi^\beta(\vec\sigma{}')\}_{P.B.}=-8i\widetilde\Phi^{\alpha\beta}\delta^p(\vec\sigma-\vec\sigma'),
\end{equation}
\begin{equation}\label{13}
\{\widetilde\Psi^\alpha(\vec\sigma),\bar{\widetilde\Psi}{}^{\dot\beta}(\vec\sigma{}')\}_{P.B.}=-4i\widetilde\Phi^{\dot\beta\alpha}\delta^p(\vec\sigma-\vec\sigma'),
\end{equation}
\begin{equation}\label{14}
\{\widetilde\Delta_W(\vec\sigma),P^{(\rho)}_M(\vec\sigma')\}_{P.B.}=2P^{(\rho)}_M\delta^p(\vec\sigma-\vec\sigma'),
\end{equation}
\begin{equation}\label{15}
\{\widetilde L_M(\vec\sigma),
P^{(\rho)}_N(\vec\sigma')\}_{P.B.}=\partial_M
P^{(\rho)}_N\delta^p(\vec\sigma-\vec\sigma'),
\end{equation}
\begin{equation}\label{16}
\{\widetilde L_M(\vec\sigma),\widetilde
L_N(\vec\sigma')\}_{P.B.}=(\widetilde
L_M(\vec\sigma')\partial_{N'}-\widetilde
L_N(\vec\sigma)\partial_M)\delta^p(\vec\sigma-\vec\sigma'),
\end{equation}
\begin{equation}\label{17}
\{\widetilde L_M(\vec\sigma),\chi(\vec\sigma')\}_{P.B.}=-\chi(\vec\sigma)\partial_M\delta^p(\vec\sigma-\vec\sigma'),
\end{equation}
 where $\chi$ are $\widetilde\Phi$, $\widetilde\Psi$ and $\widetilde\Delta_W$ constraints. The complex conjugate relations have to be added to (\ref{12})-(\ref{17}).
 The remaining P.B.'s of the  constraints are equal to zero in the strong sense. Having the algebra (\ref{12})-(\ref{17}) one can construct BRST charge of the tensionless super $p$-brane.

\section{\bf BRST charge and $OSp(1|8)$ symmetry generators}

The algebra(\ref{12})-(\ref{17}) has the rank equal unity and  may be  presented in the generalized canonical form
\begin{equation}\label{18}
\{\Upsilon^{\mathcal A}(\vec\sigma),\Upsilon^{\mathcal B}(\vec\sigma')\}_{P.B.}=\int d^p\sigma'' f^{{\mathcal A}{\mathcal B}}{}_{\mathcal C}(\vec\sigma,\vec\sigma'|\vec\sigma'')\Upsilon^{\mathcal C}(\vec\sigma''),
\end{equation}
where $ f^{{\mathcal A}{\mathcal B}}{}_{\mathcal C}$ are structure functions.
Let us note that the  algebra (\ref{18}) generalizes the original algebra \cite{BFV} by the taking into account   $\partial_M\delta^p(\vec\sigma-\vec\sigma')$ in the structure functions following from  the P.B.'s including the Virasoro constraints $\widetilde L_M(\vec\sigma)$ such as
\begin{equation}\label{19}
\begin{array}{c}
f^{\widetilde L_M\widetilde\Phi^{\dot\alpha\beta}}{}_{\widetilde\Phi^{\dot\gamma\delta}}=-\delta^{\dot\alpha}_{\dot\gamma}\delta^\beta_\delta\partial_M\delta^p(\vec\sigma-\vec\sigma')\delta^p(\vec\sigma-\vec\sigma''), \\[0.2cm]
f^{\widetilde L_M\widetilde L_N}{}_{\widetilde L_Q}= -\delta _N^Q \partial_M\delta^p(\vec\sigma-\vec\sigma')\delta^p(\vec\sigma-\vec\sigma'')
+
\delta _M^Q\partial_{N'}\delta ^p(\vec\sigma-\vec\sigma')\delta^p(\vec\sigma'-\vec\sigma'')
\end{array}
\end{equation}
and other ones.

The canonically conjugate ghost pairs of the minimal sector corresponding to the first-class constraints may be introduced forming the following triads
\begin{equation}\label{20}
\begin{array}{c}
({\widetilde\Phi}^{\alpha\beta}, C_{\alpha\beta}, \breve P^{\alpha\beta});\quad (\bar{\widetilde\Phi}{}^{\dot\alpha\dot\beta}, \bar C_{\dot\alpha\dot\beta}, \bar{\breve P}{}^{\dot\alpha\dot\beta});\quad (\widetilde\Phi^{\dot\alpha\beta}, C_{\beta\dot\alpha}, \breve P^{\dot\alpha\beta});\\
(\widetilde\Psi^\alpha, C_\alpha, \breve P^\alpha);\quad (\bar{\widetilde\Psi}{}^{\dot\alpha},\bar C_{\dot\alpha}, \bar{\breve P}{}^{\dot\alpha});\\
(\tilde P_{u}^\alpha, C_{u\alpha}, \breve P_{u}^{\alpha});\quad(\bar{\tilde P}{}^{\dot\alpha}_{u},\bar C_{u\dot\alpha}, \bar{\breve P}{}^{\dot\alpha}_{u});\\
(\tilde P^{(\rho)}_\tau, C^{(\rho)\tau}, \breve P^{(\rho)}_\tau);
\quad(P^{(\rho)}_M, C^{(\rho)M}, \breve P^{(\rho)}_M);
\\
(\widetilde L_M, C^{M}, \breve P_M);\quad(\widetilde{\Delta}_{W}, C^{(W)}, \breve P^{(W)}).\\
\end{array}
\end{equation}

Utilizing nonzero structure functions of the superalgebra (\ref{12})-(\ref{17})
 one can present the corresponding   BRST generator $\Omega$ of
 the minimal sector \cite{BFV}
\begin{equation}\label{21}
\Omega=\int d^p\sigma(C_{\mathcal A}\Upsilon^{\mathcal A}+{\textstyle\frac{1}{2}}(-)^{b}C_{\mathcal B}C_{\mathcal A}\breve f^{{\mathcal A}{\mathcal B}}{}_{\mathcal C}\breve P^{\mathcal C})(\vec\sigma),
\end{equation}
 by the  following integral along the hypersurface of the closed super $p$-brane
\begin{equation}\label{22}
\begin{array}{rl}
\Omega=\int\! d^p\sigma\!\! &\!\! \left[(C_{\alpha\beta}\widetilde\Phi^{\alpha\beta}\!+\!\bar C_{\dot\alpha\dot\beta}\bar{\widetilde\Phi}{}^{\dot\alpha\dot\beta}\!+\! C_\alpha\widetilde\Psi^\alpha\!-\!\bar C_{\dot\alpha}\bar{\widetilde\Psi}{}^{\dot\alpha}\!+\! C_{u\alpha}\widetilde P_u^\alpha\!+\!\bar C_{u\dot\alpha}\bar{\widetilde P}{}_u^{\dot\alpha})\right. \\[0.2cm]
&+C_{\alpha\dot\beta}\widetilde\Phi^{\dot\beta\alpha}\!+\! C^{(\rho)\tau}\widetilde P^{(\rho)}_\tau\!\!+\! C^{(\rho)M}P^{(\rho)}_M\!\!+\! C^{(W)}\widetilde\Delta^{ext}_W\!+\! C^{M}\widetilde L^{ext}_M\\[0.2cm]
&\left.+4i(C_\alpha C_\beta\breve P^{\alpha\beta}+\bar C_{\dot\alpha}\bar C_{\dot\beta}\bar{\breve P}{}^{\dot\alpha\dot\beta})-4iC_\alpha\bar C_{\dot\beta}\breve P^{\dot\beta\alpha}\right].\\
\end{array}
\end{equation}
 $\widetilde\Delta^{ext}_{W}$ in eq.(\ref{22}) is the generator of the gauge world-volume Weyl symmetry
\begin{equation}\label{23}
\widetilde\Delta^{ext}_{W}=\widetilde\Delta_W-2C^{(\rho)M}\breve P^{(\rho)}_M,
\end{equation}
and $\widetilde L^{ext}_M$ is the generalized Virasoro  generator
\begin{equation}\label{24}
\begin{array}{rl}
\widetilde L_M^{ext}=&\widetilde L_M+\partial_M C_{\alpha\beta}\breve P^{\alpha\beta}+
\partial_M\bar C_{\dot\alpha\dot\beta}\bar{\breve P}{}^{\dot\alpha\dot\beta}+\partial_M C_{\alpha\dot\beta}\breve P^{\dot\beta\alpha}\\
+&\partial_M C_\alpha\breve P^{\alpha}+\partial_M\bar C_{\dot\alpha}\bar{\breve P}{}^{\dot\alpha}
+\partial_M C^{(W)}\breve P^{(W)}-C^{(\rho)N}\partial_M\breve P^{(\rho)}_N
+\partial_M C^{N}\breve P_N
\end{array}
\end{equation}
extended by the ghost contributions.

Using the P.B.'s of the superalgebra (\ref{12})-(\ref{17}) one can show that the P.B. of the  BRST generator density  $\Omega(\tau,\vec\sigma)$, defined by the integrand (\ref{22}),  with itself is  equal to the total derivative
\begin{equation}\label{25}
\begin{array}{rl}
\{\Omega(\vec\sigma),\Omega(\vec\sigma')\}_{P.B.}=&\partial_M(C^M(C_{\alpha\beta}\widetilde\Phi^{\alpha\beta}+\bar C_{\dot\alpha\dot\beta}\bar{\widetilde\Phi}{}^{\dot\alpha\dot\beta}+C_{\alpha\dot\beta}\widetilde\Phi^{\dot\beta\alpha}
\\
+& C_\alpha\widetilde\Psi^{\alpha}-\bar C_{\dot\alpha}\bar{\widetilde\Psi}{}^{\dot\alpha}+C^{(\rho)N}P^{(\rho)}_N+C^{(W)}\widetilde\Delta^{ext}_W+C^N\widetilde L^{ext}_N \\
+& 4i(C_\alpha C_\beta\breve P^{\alpha\beta}+\bar C_{\dot\alpha}\bar C_{\dot\beta}\bar{\breve P}{}^{\dot\alpha\dot\beta}-C_\alpha\bar C_{\dot\beta}\breve P^{\dot\beta\alpha}))\delta^p(\vec\sigma-\vec\sigma'),
\end{array}
\end{equation}
because of the  presence of  $\partial_M\delta^p(\vec\sigma-\vec\sigma')$
in the structure functions of the superalgebra (\ref{12})-(\ref{17}).
But, the contribution of the total derivative in the r.h.s. of (\ref{25}) vanishes after integration in $\vec\sigma$ and $\vec\sigma'$ due to  the periodical boundary conditions for the closed $p$-brane. It results in the P.B.-anticommutativity of the BRST charge $\Omega\equiv\int\! d^p\sigma\Omega(\tau,\vec\sigma)$ (\ref{21}) with itself
\begin{equation}\label{26}
\{\Omega,\Omega\}_{P.B.}=0.
\end{equation}

The introduction of the ghost variables leads to the extension of the $OSp(1|8)$ symmetry generators providing the P.B.-(anti)commutativity of the $OSp(1|8)$ generators with  $\Omega$ (\ref{21}). The ghost extended "square  roots" $\widetilde S_{\gamma}(\tau,\vec\sigma)$ and  $\bar{\widetilde S}_{\dot\gamma}(\tau,\vec\sigma)$ of the  ghost extended
conformal boost densities $\widetilde K_{\gamma\dot\gamma}(\tau,\vec\sigma)$ and  $\widetilde K_{\gamma\lambda}(\tau,\vec\sigma) $ are given by
\begin{equation}\label{27}
\begin{array}{c}
\widetilde S_{\gamma}(\tau,\vec\sigma)=(z_{\gamma\delta}-2i\theta_\gamma\theta_\delta)Q^\delta +(x_{\gamma\dot\delta}-2i\theta_\gamma\bar\theta_{\dot\delta})\bar Q^{\dot\delta}
+4i(\tilde u^\delta\theta_\delta-\bar{\tilde u}^{\dot\delta}\bar\theta_{\dot\delta})\tilde{\tilde P}_{u\gamma}+\frac{2}{(\tilde\rho^\tau)^{1/2}}\tilde{\tilde P}_{u\gamma}f
\\[0.2cm]
+C_\gamma{}^\beta\breve P_\beta+C_{\gamma\dot\beta}\bar{\breve P}{}^{\dot\beta}+4iC_\gamma(\theta_\beta\breve P^{\beta}-\bar\theta_{\dot\beta}\bar{\breve P}{}^{\dot\beta})\\[0.2cm]
- 8i\theta_\delta C_\gamma{}^\beta\breve P_\beta{}^\delta+4i\bar\theta^{\dot\delta}C_\gamma{}^\beta\breve P_{\beta\dot\delta}
+4i\theta_\delta C_{\gamma\dot\beta}\breve P^{\dot\beta\delta}-8i\bar\theta^{\dot\delta} C_{\gamma\dot\beta}\bar{\breve P}{}^{\dot\beta}{}_{\dot\delta},
\end{array}
\end{equation}
\begin{equation}\label{28}
\begin{array}{c}
\bar{\widetilde S}_{\dot\gamma}(\tau,\vec\sigma)=(\bar z_{\dot\gamma\dot\delta}-2i\bar\theta_{\dot\gamma}\bar\theta_{\dot\delta})\bar Q^{\dot\delta}+(x_{\delta\dot\gamma}+2i\theta_\delta\bar\theta_{\dot\gamma})Q^{\delta}
-4i(\tilde u^\delta\theta_\delta-\bar{\tilde u}^{\dot\delta}\bar\theta_{\dot\delta})\bar{\tilde{\tilde P}}{}_{u\dot\gamma}-\frac{2}{(\tilde\rho^\tau)^{1/2}}\bar{\tilde{\tilde P}}{}_{u\dot\gamma}f
\\[0.2cm]
- C_{\beta\dot\gamma}\breve P^{\beta}-\bar C^{\dot\beta}{}_{\dot\gamma}\bar{\breve P}{}_{\dot\beta}-4i\bar C_{\dot\gamma}(\theta_\beta\breve P^{\beta}-\bar\theta_{\dot\beta}\bar{\breve P}{}^{\dot\beta})
\\[0.2cm]
- 8i\theta^\delta C_{\beta\dot\gamma}\breve P_\delta{}^\beta+4i\bar\theta_{\dot\delta}C_{\beta\dot\gamma}\breve P^{\dot\delta\beta}
+4i\theta^{\delta}\bar C^{\dot\beta}{}_{\dot\gamma}\breve P_{\delta\dot\beta}-8i\bar\theta_{\dot\delta}\bar C^{\dot\beta}{}_{\dot\gamma}\bar{\breve P}{}^{\dot\delta}{}_{\dot\beta}.
\end{array}
\end{equation}
Using the densities $ \Omega(\tau,\vec\sigma)$ (\ref{22}) and  $ \widetilde S_{\gamma}(\tau,\vec\sigma')$   (\ref{27})  we find their P.B.
\begin{equation}\label{29}
\{\Omega(\sigma),\widetilde S_{\gamma}(\vec\sigma')\}_{P.B.}=-(C^M \widetilde S_{\gamma})(\vec\sigma)\partial_M\delta^p(\sigma-\sigma')
\end{equation}
and conclude that the contribution of the total derivative in the r.h.s. of (\ref{29}) vanishes after integration with respect to $\vec\sigma$ and $\vec\sigma'$.
Thus, the BRST charge  $\Omega$ (\ref{22}) has zero P.B.'s with the conformal supercharges $\widetilde S_{\gamma},\bar{\widetilde S}_{\dot\gamma} $
\begin{equation}\label{30}
\{\Omega,\widetilde S_{\gamma}\}_{P.B.}=0,\quad \{\Omega,\bar{\widetilde S}_{\dot\gamma}\}_{P.B.}=0.
\end{equation}
The same P.B.-commutativity
 \begin{equation}\label{31}
\{\Omega,G\}_{P.B.}=0
\end{equation}
  between  $\Omega$ and other $OSp(1|8)$ symmetry charges
$G\equiv \int d^p\sigma G(\tau,\vec\sigma)$ extended by the  ghost contributions will also be preserved, because of the general relation (\ref{17}) for the generator densities:
$\{\widetilde L_M(\vec\sigma),G(\vec\sigma')\}_{P.B.}=-G(\vec\sigma)\partial_M\delta(\vec\sigma-\vec\sigma')$.

The above mentioned expressions for the generator densities of the
generalized conformal transformations extended by the ghost contributions take the form\footnote{The discussed formulae would look more compact in the Majorana spinor representation. However, it makes more obscure the contribution of the TCC coordinates $z_{\alpha\beta}$ which presence is crucial for the exotic supersymmetry protection and generation of the new bosonic gauge symmetries (see \cite{ZUB}) generalizing the well known symmetries of the Penrose twistor approach originally formulated in $D=4$ \cite{PR}. In the Majorana representation $z_{\alpha\beta}$ are encoded in the symmetric $4\times4$ matrix $Y_a{}^b\equiv Y_{ad}C^{db}=\left(\begin{array}{cc} z_\alpha{}^\beta & x_{\alpha\dot\beta}\\ \tilde x^{\dot\alpha\beta} & \bar z^{\dot\alpha}{}_{\dot\beta}\end{array}\right)$ together with the space-time coordinates $x_{\alpha\dot\alpha}$ (see \cite{ZU}, \cite{UZ03}).}
\begin{equation}\label{32}
\begin{array}{c}
{\widetilde K}_{\gamma\lambda}(\tau,\vec\sigma)= 2z_{\gamma\beta}z_{\lambda\delta}\pi^{\beta\delta}+2x_{\gamma\dot\beta}x_{\lambda\dot\delta}\bar\pi^{\dot\beta\dot\delta}+z_{\gamma\beta}x_{\lambda\dot\delta}P^{\dot\delta\beta}+x_{\gamma\dot\beta}z_{\lambda\delta}
P^{\dot\beta\delta}
\\[0.2cm]
+ \theta_\lambda(z_{\gamma\delta}\pi^{\delta}+x_{\gamma\dot\delta}\bar\pi^{\dot\delta})+\theta_{\gamma}(z_{\lambda\delta}\pi^{\delta}+x_{\lambda\dot\beta}\bar\pi^{\dot\delta})
+ (\tilde u^{\delta} z_{\delta\lambda}-\bar{\tilde u}{}^{\dot\delta} x_{\lambda\dot\delta})\tilde{\tilde P}_{
u\gamma}+(\tilde u^\delta z_{\delta\gamma}-\bar{\tilde u}{}^{\dot\delta}x_{\gamma\dot\delta})\tilde{\tilde P}_{u\lambda}
\\[0.2cm]
- 2i(\tilde u^\delta\theta_\delta-\bar{\tilde u}{}^{\dot\delta}\bar\theta_{\dot\delta})(\theta_{\lambda}\tilde{\tilde P}_{u\gamma}+\theta_\gamma\tilde{\tilde P}_{u\lambda})
+{\textstyle\frac{2}{(\tilde\rho^\tau)^{1/2}}}(\theta_{\lambda}\tilde{\tilde P}_{u\gamma}+\theta_\gamma\tilde{\tilde P}_{u\lambda})f-{\textstyle\frac{1}{\tilde\rho^\tau}}\tilde{\tilde P}_{u\gamma}\tilde{\tilde P}_{u\lambda}
\\[0.2cm]
+\theta_\lambda C_\gamma{}^\beta\breve P_\beta+\theta_\gamma C_\lambda{}^\beta\breve P_\beta
+\theta_\lambda C_{\gamma\dot\beta}\bar{\breve P}{}^{\dot\beta}+\theta_\gamma C_{\lambda\dot\beta}\bar{\breve P}{}^{\dot\beta}
\\[0.2cm]
-(z_\lambda{}^\beta+2i\theta_\lambda\theta^\beta)C_\gamma\breve P_\beta-(z_\gamma{}^\beta+2i\theta_\gamma\theta^\beta)C_\lambda\breve P_\beta
-(x_{\lambda\dot\beta}+2i\theta_\lambda\bar\theta_{\dot\beta})C_\gamma\bar{\breve P}{}^{\dot\beta}-(x_{\gamma\dot\beta}+2i\theta_\gamma\bar\theta_{\dot\beta})C_\lambda\bar{\breve P}{}^{\dot\beta}
\\[0.2cm]
-2(z_\gamma{}^\beta+2i\theta_\gamma\theta^\beta)C_{\lambda\delta}\breve P_\beta{}^\delta\!-\!2(z_\lambda{}^\beta+2i\theta_\lambda\theta^\beta)C_{\gamma\delta}\breve P_\beta{}^\delta
\\[0.2cm]
\!+\!(z_{\gamma\beta}+2i\theta_\gamma\theta_\beta)C_{\lambda\dot\delta}\breve P^{\dot\delta\beta}\!+\!(z_{\lambda\beta}+2i\theta_\lambda\theta_\beta)C_{\gamma\dot\delta}\breve P^{\dot\delta\beta}\\[0.2cm]
+(x_{\gamma\dot\beta}+2i\theta_\gamma\bar\theta_{\dot\beta})C_{\lambda\delta}\breve P^{\dot\beta\delta}\!\!+\!(x_{\lambda\dot\beta}+2i\theta_\lambda\bar\theta_{\dot\beta})C_{\gamma\delta}\breve P^{\dot\beta\delta}
\\[0.2cm]
\!\!+2(x_{\gamma\dot\beta}+2i\theta_\gamma\bar\theta_{\dot\beta})C_{\lambda\dot\delta}\bar{\breve P}{}^{\dot\beta\dot\delta}\!\!+\!(x_{\lambda\dot\beta}+2i\theta_\lambda\bar\theta_{\dot\beta})C_{\gamma\dot\delta}\bar{\breve P}{}^{\dot\beta\dot\delta},
\end{array}
\end{equation}
 for  ${\widetilde K}_{\gamma\lambda}(\tau,\vec\sigma)$  and respectively  for
 ${\widetilde K}_{\gamma\dot\gamma}(\tau,\vec\sigma)$
\begin{equation}\label{33}
\begin{array}{c}
{\widetilde K}_{\gamma\dot\gamma}(\tau,\vec\sigma)= z_{\gamma\delta}\bar z_{\dot\gamma\dot\delta}P^{\dot\delta\delta}+x_{\gamma\dot\delta}x_{\delta\dot\gamma}P^{\dot\delta\delta}+2(z_{\gamma\delta}x_{\lambda\dot\gamma}\pi^{\delta\lambda}+x_{\gamma\dot\delta}\bar z_{\dot\gamma\dot\lambda}\bar\pi^{\dot\delta\dot\lambda})
\\[0.2cm]
+ \theta_\gamma(\bar z_{\dot\gamma\dot\delta}\bar\pi^{\dot\delta}+x_{\delta\dot\gamma}\pi^{\delta})+\bar\theta_{\dot\gamma}(z_{\gamma\delta}\pi^{\delta}+x_{\gamma\dot\delta}\bar\pi^{\dot\delta})
+ (\tilde u^{\delta} x_{\delta\dot\gamma}-\bar{\tilde u}{}^{\dot\delta}\bar z_{\dot\delta\dot\gamma})\tilde{\tilde P}_{u\gamma}+(x_{\gamma\dot\delta}\bar{\tilde u}{}^{\dot\delta}-z_{\gamma\delta}\tilde u^\delta)\bar{\tilde{\tilde P}}_{u\dot\gamma}
\\[0.2cm]
- 2i(\tilde u^\delta\theta_\delta-\bar{\tilde u}{}^{\dot\delta}\bar\theta_{\dot\delta})(\bar\theta_{\dot\gamma}\tilde{\tilde P}_{u\gamma}-\theta_\gamma \bar{\tilde{\tilde P}}_{u\dot\gamma})
+{\textstyle\frac{2}{(\tilde\rho^\tau)^{1/2}}}(\bar\theta_{\dot\gamma}\tilde{\tilde P}_{u\gamma}-\theta_\gamma \bar{\tilde{\tilde P}}_{u\dot\gamma})f+{\textstyle\frac{1}{\tilde\rho^\tau}}\tilde{\tilde P}_{u\gamma}\bar{\tilde{\tilde P}}_{u\dot\gamma}
\\[0.2cm]
+\bar\theta_{\dot\gamma}C_\gamma{}^\beta\breve P_{\beta}-\theta_\gamma\bar C_{\dot\gamma}{}^{\dot\beta}\bar{\breve P}_{\dot\beta}
+\bar\theta_{\dot\gamma}C_{\gamma\dot\beta}\bar{\breve P}{}^{\dot\beta}-\theta_\gamma C_{\beta\dot\gamma}\breve P^{\beta}
\\[0.2cm]
+(\bar z_{\dot\gamma}{}^{\dot\beta}+2i\bar\theta_{\dot\gamma}\bar\theta^{\dot\beta})C_\gamma\bar{\breve P}_{\dot\beta}+(z_\gamma{}^\beta+2i\theta_\gamma\theta^\beta)\bar C_{\dot\gamma}\breve P_\beta
+(x_{\beta\dot\gamma}-2i\theta_\beta\bar\theta_{\dot\gamma})C_\gamma\breve P^{\beta}+(x_{\gamma\dot\beta}+2i\theta_\gamma\bar\theta_{\dot\beta})\bar C_{\dot\gamma}\bar{\breve P}{}^{\dot\beta}
\\[0.2cm]
-2(z_\gamma{}^\beta+2i\theta_\gamma\theta^\beta)C_{\delta\dot\gamma}\breve P_\beta{}^\delta\!\!-\!2(\bar z_{\dot\gamma}{}^{\dot\beta}+2i\bar\theta_{\dot\gamma}\bar\theta^{\dot\beta})\bar C_{\gamma\dot\delta}\bar{\breve P}{}^{\dot\delta}{}_{\dot\beta}
\\[0.2cm]
\!\!+\!(z_\gamma{}^\beta+2i\theta_\gamma\theta^\beta)\bar C^{\dot\delta}{}_{\dot\gamma}\breve P_{\beta\dot\delta}\!\!+\!(\bar z^{\dot\beta}{}_{\dot\gamma}+2i\bar\theta_{\dot\gamma}\bar\theta^{\dot\beta})C_{\gamma}{}^\delta\bar{\breve P}_{\delta\dot\beta}
\\[0.2cm]
+(x_{\gamma\dot\beta}+2i\theta_\gamma\bar\theta_{\dot\beta})C_{\delta\dot\gamma}\breve P^{\dot\beta\delta}+(x_{\beta\dot\gamma}-2i\theta_\beta\bar\theta_{\dot\gamma})C_{\gamma\dot\delta}\breve P^{\dot\delta\beta}
\\[0.2cm]
-2(x_{\gamma\dot\beta}+2i\theta_\gamma\bar\theta_{\dot\beta})\bar C^{\dot\delta}{}_{\dot\gamma}\bar{\breve P}{}^{\dot\beta}{}_{\dot\delta}-2(x_{\beta\dot\gamma}-2i\theta_\beta\bar\theta_{\dot\gamma})C_\gamma{}^\delta\breve P_{\delta}{}^\beta.\\
\end{array}
\end{equation}
The remaining 16 generator  densities of $OSp(1|8)$ supergroup extended by the ghosts are the following
\begin{equation}
\begin{array}{c}\label{34}
{\widetilde L}^{\alpha}{}_{\beta}(\tau,\vec\sigma)=x_{\beta\dot\beta}P^{\dot\beta\alpha}+
2z_{\gamma\beta}\pi^{\alpha\gamma}+\theta_\beta\pi^\alpha +\tilde u^\alpha\tilde{\tilde P}_{u\beta}
-2C_\beta{}^\gamma\breve P_\gamma{}^\alpha+C_{\beta\dot\gamma}\breve P^{\dot\gamma\alpha}+C_\beta\breve P^{\alpha},\\
{\widetilde L}^\alpha{}_{\dot\beta}(\tau,\vec\sigma)=2x_{\gamma\dot\beta}\pi^{\alpha\gamma} +
\bar z_{\dot\beta\dot\gamma}P^{\dot\gamma\alpha}
+\bar\theta_{\dot\beta}\pi^\alpha -\tilde u^\alpha \bar{\tilde{\tilde P}}_{u\dot\beta}
+2C_{\gamma\dot\beta}\breve  P^{\gamma\alpha}+\bar C_{\dot\beta\dot\gamma}\breve P^{\dot\gamma\alpha}-\bar C_{\dot\beta}\breve P^{\alpha}.
\end{array}
\end{equation}
The adduced expressions should be complemented by their complex
conjugate.

Note that supersymmetry and generalized translation
generator densities do not contain  any ghost contribution. One  can
check that the P.B.-commutation relations of the  $OSp(1|8)$
superalgebra  extended by the  ghost contributions coincide  with
the P.B.-commutation relations of the original $OSp(1|8)$
superalgebra \cite{UZ03}.

\section{\bf Quantization: nilpotent BRST operator and quantum $OSp(1|8)$ algebra}
Upon transition to quantum theory  all the quantities entering the
converted constraints
and $OSp(1|8)$ generator densities
are treated as operators that implies a choice of certain ordering for products of noncommuting operators. At the same time  the canonical Poisson brackets
$
\{{\cal P}^{\mathfrak M}(\vec\sigma),{\cal Q}_{\mathfrak
N}(\vec\sigma')\}_{P.B.}=\delta^{\mathfrak M}_{\mathfrak
N}\delta^p(\vec\sigma-\vec\sigma ')
$
 used in \cite{UZ03} transform into (anti)commutators
\begin{equation}\label{35}
[{\hat{\cal P}}^{\mathfrak M}(\vec\sigma),{\hat{\cal Q}}_{\mathfrak
N}(\vec\sigma')\}=-i\delta^{\mathfrak M}_{\mathfrak
N}\delta^p(\vec\sigma-\vec\sigma').
\end{equation}
 It is necessary to provide  further nilpotence of the BRST operator, fulfilment of (anti)commutation relations of the
$OSp(1|8)$ superalgebra and its generator (anti)\-com\-mu\-ta\-ti\-vi\-ty with  the BRST operator ensuring the global quantum invariance of the model.
In addition, the Hermiticity of the quantum  BRST operator and $OSp(1|8)$ generators has to be supported.
The Hermiticity requirement may be manifestly  satisfied if we start from the above constructed classical representations  for the  $OSp(1|8)$ generators and BRST charge in which  all coordinates  are  disposed  from the left of momenta, i.e. in the form $\cal Q\cal P$, where $\cal Q$ and $\cal P$ are the products of the coordinates and momenta contained in $\Omega$ and the  generators.
 Then the operator expressions for the latter are presented  in the manifestly Hermitian form composed of the operator products
 $\frac12(\hat{\cal Q}\hat{\cal P}+(-)^{\epsilon({\cal Q})\epsilon({\cal P})}(\hat{\cal Q}\hat{\cal P})^\dagger)$,
where $\epsilon({\cal Q})$ and $\epsilon({\cal P})$ are Grassmannian gradings of these coordinate and momentum monomials.

In particular,  we obtain  the following  Hermitian operator
representations
\begin{equation}\label{36}\begin{array}{c}
\hat{\widetilde{\mathcal S}}_\gamma(\tau,\vec\sigma)={\textstyle\frac12}(\hat z_{\gamma\delta}-2i\hat\theta_\gamma\hat\theta_\delta)\hat Q^\delta+{\textstyle\frac12}\hat Q^\delta (\hat z_{\gamma\delta}
-2i\hat\theta_\gamma\hat\theta_\delta)
\\[0.2cm]
+{\textstyle\frac12}(\hat x_{\gamma\dot\delta}-2i\hat\theta_\gamma\hat{\bar\theta}_{\dot\delta})\hat{\bar Q}{}^{\dot\delta}+{\textstyle\frac12}\hat{\bar Q}{}^{\dot\delta}(\hat x_{\gamma\dot\delta}-2i\hat\theta_\gamma\hat{\bar\theta}_{\dot\delta})
\\[0.2cm]
+2i\hat\theta_\delta(\hat{\tilde u}{}^\delta\hat{\tilde{\tilde P}}_{u\gamma}+\hat{\tilde{\tilde P}}_{u\gamma}\hat{\tilde u}{}^\delta)-4i\hat{\bar{\tilde u}}{}^{\dot\delta}\hat{\bar\theta}_{\dot\delta}\hat{\tilde{\tilde P}}_{u\gamma}+{\textstyle\frac{2}{(\hat{\tilde\rho}{}^{\tau})^{1/2}}}\hat f\hat{\tilde{\tilde P}}_{u\gamma}
\\[0.2cm]
+\hat C_\gamma{}^\beta\hat{\breve P}_\beta+\hat C_{\gamma\dot\beta}\hat{\bar{\breve P}}{}^{\dot\beta}
+2i\hat\theta_\delta(\hat C_\gamma\hat{\breve P}{}^{\delta}+\hat{\breve P}{}^{\delta}\hat C_\gamma)+4i\hat{\bar\theta}{}^{\dot\delta}\hat C_\gamma\hat{\bar{\breve P}}_{\dot\delta}
\\[0.2cm]
-4i\hat\theta_\delta(\hat C_\gamma{}^\beta\hat{\breve P}_\beta{}^\delta-\hat{\breve P}_\beta{}^\delta \hat C_\gamma{}^\beta)+4i\hat{\bar\theta}{}^{\dot\delta}\hat C_\gamma{}^\beta\hat{\breve P}_{\beta\dot\delta}
+2i\hat\theta_\delta(\hat C_{\gamma\dot\beta}\hat{\breve P}{}^{\dot\beta\delta}-\hat{\breve P}{}^{\dot\beta\delta}\hat C_{\gamma\dot\beta})-8i\hat{\bar\theta}{}^{\dot\delta}\hat C_{\gamma\dot\beta}\hat{\bar{\breve P}}{}^{\dot\beta}{}_{\dot\delta}
\end{array}
\end{equation}
for the classical density  $\widetilde S_{\gamma}(\tau,\vec\sigma)$ (\ref{27})
and
\begin{equation}\label{37}
\begin{array}{c}
\hat{\widetilde{\mathcal L}}{}^\alpha{}_\beta(\tau,\vec\sigma)=\frac12(\hat x_{\beta\dot\beta}\hat P^{\dot\beta\alpha}+\hat P^{\dot\beta\alpha}\hat x_{\beta\dot\beta})+(\hat z_{\beta\gamma}\hat\pi^{\gamma\alpha}+\hat\pi^{\gamma\alpha}\hat z_{\beta\gamma})
\\[0.2cm]
+\frac12(\hat\theta_\beta\hat\pi^\alpha-\hat\pi^\alpha\hat\theta_\beta)+\frac12(\hat{\tilde u}{}^\alpha\hat{\tilde{\tilde P}}_{u\beta}+\hat{\tilde{\tilde P}}_{u\beta}\hat{\tilde u}{}^\alpha)\\[0.2cm]
-(\hat C_\beta{}^\gamma\hat{\breve P}_\gamma{}^\alpha-\hat{\breve P}_\gamma{}^\alpha\hat C_\beta{}^\gamma)+\frac12(\hat C_{\beta\dot\gamma}\hat{\breve P}{}^{\dot\gamma\alpha}-\hat{\breve P}{}^{\dot\gamma\alpha}\hat C_{\beta\dot\gamma})
+\frac12(\hat C_\beta\hat{\breve P}{}^{\alpha}+\hat{\breve P}{}^{\alpha}\hat C_\beta),
\end{array}
\end{equation}
 for the generalized Lorentz density  ${\widetilde L}{}^\alpha{}_\beta(\tau,\vec\sigma)$ (\ref{34}).

By the same  way  can be constructed quantum Hermitian
 generators $\hat{\widetilde L}{}^{ext}_{M} $ of the $\vec\sigma$-reparametrizations
 and the  Weyl symmetry generator $\hat{\widetilde\Delta}{}^{ext}_W$
\begin{equation}\label{38}
\begin{array}{c}
\hat{\widetilde\Delta}{}^{ext}_W={\textstyle\frac12}(\hat{\tilde
u}_\alpha\hat{\tilde{\tilde P}}{}^\alpha_u+\hat{\tilde{\tilde P}}{}^\alpha_u\hat{\tilde
u}_\alpha+\hat{\bar{\tilde u}}_{\dot\alpha}\hat{\bar{\tilde{\tilde P}}}{}^{\dot\alpha}_u
+\hat{\bar{\tilde{\tilde P}}}{}^{\dot\alpha}_u\hat{\bar{\tilde u}}_{\dot\alpha})\\[0.2cm]
-\hat{\tilde\rho}{}^\tau\hat{\tilde{\tilde
P}}{}^{(\rho)}_\tau-\hat{\tilde{\tilde
P}}{}^{(\rho)}_\tau\hat{\tilde\rho}{}^\tau-\hat\rho^M\hat P^{(\rho)}_M-\hat P^{(\rho)}_M\hat\rho^M-\hat C^{(\rho)M}\hat{\breve P}{}^{(\rho)}_M+\hat{\breve P}{}^{(\rho)}_M\hat C^{(\rho)M}\approx0,
\end{array}
\end{equation}
 entering the BRST operator.

 Because other converted first-class constraints are Hermitian by construction,
the quantum Hermitian BRST generator
\begin{equation}\label{39'}
\hat{\bf\Omega}=\frac12\int d^p\sigma(\hat\Omega(\tau,\vec\sigma)+\hat\Omega^\dagger(\tau,\vec\sigma))
\end{equation}
will coincide with its classical expression $\Omega$ (\ref{22}) after the substitution  of (\ref{38}) and the Hermitian representation for $\hat{\widetilde L}_M(\vec\sigma)$  originated from (\ref{24}) in eq. (\ref{22}).

Now we are ready to prove that this realization of
$\hat{\bf\Omega}$ preserves its nilpotency and (anti)commutativity
with the Hermitian operators (\ref{36}), (\ref{37}) and other ones
generating a quantum realization of the  classical $OSp(1|8)$
superalgebra. The proof is obvious and is  based on the observation
that $\hat{\bf\Omega}$ and other considered Hermitian operators
are linear in the momentum operators $\hat{\cal P}^{\mathfrak
M}(\tau,\vec\sigma)$ of the original coordinates and ghost fields.
The remarkable property of the ordered polynomial operators
composed of $\hat{\cal Q}_{\mathfrak M}(\tau,\vec\sigma)$ and
$\hat{\cal P}^{\mathfrak M}(\tau,\vec\sigma)$, which form the
Weyl-Heisenberg algebra (\ref{35}), and are linear in $\hat{\cal
P}^{\mathfrak M}$ is the preservation of the chosen ordering in
course of calculations of their (anti)commutators. As a result,
the transition from the P.B.'s to (anti)commutators will preserve
all classical results obtained in the P.B. realization of the
extended algebra of the $OSp(1|8)$ generators and classical BRST
charge of the super $p$-brane. So, the quantum Hermitian BRST
operator (\ref{39'}) occurs to be nilpotent
\begin{equation}\label{40'}
\{\hat{\bf\Omega},\hat{\bf\Omega}\}=0.
\end{equation}

However, the Hermiticity of  $\hat{\bf\Omega}$ and the $OSp(1|8)$
generating operators by itself is only a necessary condition for
the quantum realization of the  physical operators, because the
relevant vacuum and physical states have also to be constructed.
So, the problem of existence of the selfconsistent quantum
realization of the exotic BPS states by the states of quantum
tensionless super $p$-brane is reduced to the proof of existence of
the relevant vacuum and the corresponding physical space of
quantum states. At the present time we investigate this problem. However, we should like to discuss here a possible way to solve this problem based on the consideration of the $\hat{\cal Q}\hat{\cal P}$-ordering studied in \cite{GLSSU}.

\section{\bf $\hat{\cal Q}\hat{\cal P}$-ordering and regularization}

 It was motivated in
 \cite {GLSSU} that the  coordinate  and  momenta operators have to be used for the tensionless  string quantization instead of the  creation and  annihilation operators  relevant  for  the tensile string  quantization.
This  motivation is physically justified by the absence of oscillator excitations for the tensionless string which makes its dynamics resembling that of collection of free particles and results in the choice of the physical vacuum as a state
 annihilated by the string  momentum operator.
 In that  case the coordinate  $\hat{\cal Q}$ and
momentum $\hat{\cal P}$  monomials
forming the above discussed  Hermitian operators have to be ordered by the shifts of  all the  $\hat {\cal Q}$ monomials  to the left of  $\hat{\cal P}$.
To achieve  that $\hat{\cal Q}\hat{\cal P}$-ordering we have, in particular, to permutate some noncommuting coordinate and momentum  operators in  the $\hat{\cal Q}\hat{\cal P}$-disordered Hermitian expressions of  the generator densities (\ref{36}), (\ref{37}) and others, constraints  $\hat{\widetilde\Delta}{}^{ext}_W$ (\ref{38}), $\hat{\widetilde L}{}^{ext}_{M} $ and  the Hermitian BRST operator.
In view of that permutations divergent terms will appear in some monomials composed from  canonically conjugate operators at coinciding  points of $p$-brane.
 A typical form of such divergent terms is illustrated by the relation
\begin{equation}\label{39}
\frac12(\hat{z}_{\lambda\delta}(\vec\sigma)\hat{\pi}^{\delta\varepsilon}(\vec\sigma)+\hat{\pi}^{\delta\varepsilon}(\vec\sigma)\hat{z}_{\lambda\delta}(\vec\sigma))=\hat{z}_{\lambda\delta}(\vec\sigma)\hat{\pi}^{\delta\varepsilon}(\vec\sigma)-\frac{3i}{4}\delta_\lambda^\varepsilon\delta^p(0)
\end{equation}
encoding the permutation effect of the TCC coordinates with their momenta.
The r.h.s. of ambiguously defined relation (\ref{39}) includes the divergent term $\delta^p(0)=\delta^p(\vec\sigma-\vec\sigma')|{}_{\vec\sigma=\vec\sigma'}$ and the problem appears how to deal with the $\hat{\cal Q}\hat{\cal P}$-ordered representation of the symmetric operator in the l.h.s. of (\ref{39}). Such type a problem is typical in quantum field theory due to its inherit divergencies and the regularization procedure should be applied. Thus, in general, the ordering problem has to be analyzed together with the divergency problem. It might have happened that a regularization prescribes to use the image of the delta function at the zero point (here $\delta^p(0)$) to be equal to zero. Then the choice of that regularization would solve the ordering problem. Taking into account such a possibility requires study of the structure of the divergent terms appearing in the total quantum algebra of our model. To this end we firstly analyze the $\hat{\cal Q}\hat{\cal P}$-ordered realization of the $OSp(1|8)$ algebra.

We find that the application of the described $\hat{\mathcal Q}\hat{\mathcal P}$-ordering procedure to the Hermitian $OSp(1|8)$ generator densities (\ref{36}), (\ref{37}) and others yields the relations
\begin{equation}\label{41}
\begin{array}{c}
\hat{\widetilde{\mathcal S}}_{\gamma}(\tau,\vec\sigma)=\hat{\widetilde S}_{\gamma}-
2\hat\theta_\gamma\delta^p(0),\
\hat{\bar{\widetilde{\mathcal S}}}_{\dot\gamma}(\tau,\vec\sigma)=\hat{\bar{\widetilde S}}_{\dot\gamma}-2\hat{\bar{\theta}}_{\dot\gamma}\delta^p(0),\\
\hat{\widetilde{\mathcal K}}_{\gamma\lambda}(\tau,\vec\sigma)=\hat
{\widetilde K}_{\gamma\lambda}+i\hat z_{\gamma\lambda}\delta^p(0),\
 \hat{\bar{\widetilde{\mathcal K}}}_{\dot\gamma\dot\lambda}(\tau,\vec\sigma)=\hat{\bar {\widetilde K}}_{\dot\gamma\dot\lambda}+i\hat{\bar z}_{\dot\gamma\dot\lambda}\delta^p(0),\\
\hat{\widetilde{\mathcal K}}_{\gamma\dot\gamma}(\tau,\vec\sigma)=\hat {\widetilde K}_{\gamma\dot\gamma}+i\hat x_{\gamma\dot\gamma}\delta^p(0),\\
\hat{\widetilde{\mathcal L}}_{\alpha\beta}(\tau,\vec\sigma)=\hat {\widetilde L}_{\alpha\beta}+\frac{i}{2}\varepsilon_{\alpha\beta}\delta^p(0),\
\hat{\bar{\widetilde{\mathcal L}}}_{\dot\alpha\dot\beta}(\tau,\vec\sigma)=\hat{\bar{\widetilde L}}_{\dot\alpha\dot\beta}+\frac{i}{2}\bar{\varepsilon}_{\dot\alpha\dot\beta}\delta^p(0)
\end{array}
\end{equation}
connecting the Hermitian and $\hat{\mathcal Q}\hat{\mathcal P}$-ordered representations for the generator densities.
The operators  $\hat{\widetilde S}_{\gamma}$, $\hat{\bar{\widetilde S}}_{\dot\gamma}$, $\hat{\widetilde{K}}_{\gamma\lambda}$, $\hat{\widetilde{K}}_{\dot\gamma\dot\lambda}$, $\hat{\widetilde{K}}_{\gamma\dot\gamma}$, $\hat{\widetilde{L}}_{\alpha\beta}$, $\hat{\bar{\widetilde L}}_{\dot\alpha\dot\beta}$, collectively denoted by $\hat{\widetilde G}_{qp}$, in the r.h.s. of (\ref{41})
 coincide with the  classical  $\hat{\mathcal Q}\hat{\mathcal P}$-ordered
representations (\ref{27}), (\ref{28}), (\ref{32})-(\ref{34}),
where the corresponding  operators are substituted for the
classical  coordinates and momenta. These $\hat{\cal Q}\hat{\cal P}$-ordered operators $\hat{\widetilde G}_{qp}$ form another representation of the quantum algebra $OSp(1|8)$ similarly to the Hermitian operators, because they originate from the classical generators and preserve the $\hat{\cal Q}\hat{\cal P}$-ordering in the course of the calculation of their
(anti)commutators. But, the $\hat{\cal Q}\hat{\cal P}$-ordered generators $\hat{\widetilde G}_{qp}$ are nonHermitian and their (anti)commutation with the divergent nonHermitian terms in the r.h.s. of the representation (\ref{41}) contributes to the closure of the $OSp(1|8)$ algebra presented by the Hermitian generators in the l.h.s. of (\ref{41}). So, we obtain two quantum realizations of the $OSp(1|8)$ algebra and one of them $\hat{\widetilde G}_{qp}$ is nonHermitian, because of its  $\hat{\cal Q}\hat{\cal P}$-ordering. But, namely, action of the nonHermitian generators $\hat{\widetilde G}_{qp}$ on the corresponding vacuum state is well defined in accordance with \cite{GLSSU}. Then the regularization assumption, which prescribes to accept the regularized image of $\delta^p(0)$ as equal zero, removes the nonHermiticity of the $\hat{\cal Q}\hat{\cal P}$-ordered generators $\hat{\widetilde G}_{qp}$. In the result we obtain the desired vacuum state for the discussed Hermitian  realization of the physical operators. So, the choice of such a regularization would allow to overcome the problem of construction of the quantum space of physical states. To realize such a scenario one needs to analyze the $\hat{\mathcal Q}\hat{\mathcal P}$-ordered realization of Hermitian BRST generator (\ref{39'}).

The correspondent $\hat{\mathcal Q}\hat{\mathcal P}$-ordered BRST operator $\hat\Omega$ turns out to be nonHermitian, as follows from the relation connecting $\hat\Omega$ with the Hermitian BRST operator $\hat{\bf\Omega}$ (\ref{39'})
\begin{equation}\label{42}
\hat{\bf\Omega}=\hat\Omega-i\int d^p\sigma
[\hat C^{(W)}
+
\textstyle{(\frac{p}{2}-\frac74)}\hat C^M\partial_M
+
\textstyle{\frac12}\partial_M\hat C^M]\delta^p(0),
\end{equation}
and is supplemented by three antiHermitian divergent additions in the r.h.s. compensating the nonHermiticity of $\hat\Omega$. The first of them, proportional to $\delta^p(0)$, follows  from the $\hat{\mathcal Q}\hat{\mathcal P}$-ordering of the Weyl symmetry generator $\hat{\widetilde\Delta}{}^{ext}_W$ (\ref{38}) and is contributed only by the auxiliary variables
$\hat{\tilde\rho}{}^{\tau} ,\hat{\tilde u}{}^{\alpha}$ and $ \hat{\bar{\tilde u}}{}^{\dot\alpha}$ partially cancelling each other during the ordering with their momenta. The contribution of other auxiliary pair $(\hat\rho^{M},\hat P^{(\rho)}_M ) $ in here is cancelled by the ghost pair  $(\hat C^{(\rho)M}, \hat{\breve P}{}^{(\rho)}_M)$ contribution.
We observe that only the auxiliary twistor-like fields and the component $\hat{\tilde\rho}{}^{\tau}$ of the world-volume density $\hat\rho^{\mu}$, introduced to provide the Weyl and reparametrization gauge symmetries of the brane action,  contributed to the first singular term. The similar  story concerns
the  second divergent term proportional to $\partial_M\delta^p(0)$
\footnote{In the symmetric regularization, where  $\delta^p_\epsilon(-\vec\sigma)=\delta^p_\epsilon(\vec\sigma)$, the derivative $\partial_M\delta^p_\epsilon(\vec\sigma)$ vanishes at $\vec\sigma=0$. As a result the symmetric regularization does not capture the divergencies
 following from the reordering of terms with derivatives like $\hat{\pi}^{\delta\varepsilon}(\vec\sigma)\partial_{M}\hat{z}_{\lambda\delta}(\vec\sigma)$ 
(cf. (\ref{39})). To capture such a type singularity one can use more general regularization of delta function considered by H\" ormander \cite {Her}. As a result, we find the r.h.s. in the regularized (anti)commutators
$[{\hat{\cal P}}^{\mathfrak M}(\vec\sigma),
\partial_{M}{\hat{\cal Q}}_{\mathfrak N}(\vec\sigma)\}=i
\delta^{\mathfrak M}_{\mathfrak N}
\partial_{M}\delta^{p}(\vec\sigma,\vec\sigma)$, where $\partial_{M}\delta^{p}(\vec\sigma,\vec\sigma)$ is a regularized analogue of $\partial_M\delta^p(0)$, to be non zero.}. This term appears from the
$\hat{\mathcal Q}\hat{\mathcal P}$-ordering of the extended Virasoro operators  $\hat{\widetilde L}{}^{ext}_{M} $ and only the above mentioned auxiliary fields  together with the ghost pairs
$({\hat C}^{M}, \hat{\breve P}_M)$, $ (\hat C^{(W)},
\hat{\breve P}{}^{(W)})$ and the auxiliary  fermionic  field
$\hat f$ contribute to here, because  the contributions of the propagating phase-space  variables are  cancelled by the corresponding  ghosts. This cancellation illustrates  the  boson-fermion  cancellation mechanism  provided by the BRST symmetry.
The third  singular addition, proportional to the  total derivative, restores Hermiticity of the cubic term $ {\hat C}^M\partial_M \hat C^{N}\hat{\breve P}_{N}$ but, it vanishes in view of the periodical boundary conditions. The latter could contribute in the case on a nontrivial topology of the ghost-field space.

Now let us note that the $\hat{\mathcal Q}\hat{\mathcal P}$-ordered operator $\hat\Omega$ in the r.h.s. of eq.(\ref{42}) is nilpotent
\begin{equation}\label{44'}
\{\hat\Omega,\hat\Omega\}=0,
\end{equation}
because of the nilpotence of the classical BRST charge $\Omega$ and its linearity in momenta of $\hat\Omega$. Thus, the singular representation (\ref{42}) contains two nilpotent operators $\hat{\bf\Omega}$ and $\hat\Omega$ in view of (\ref{40'}) and (\ref{44'}). Then, the calculation of the anticommutators of the l.h.s. of eq.(\ref{42}) with itself and the r.h.s. with itself results in the condition
\begin{equation}\label{43}
\{\hat{\bf\Omega},\hat{\bf\Omega}\}=0=-\int d^p\sigma[2\hat C^{M}\partial_M\hat C^{(W)}+
\textstyle{(p-\frac72)}\hat C^M\partial_M\hat C^N\partial_N]\delta^p(0).
\end{equation}

To clarify the expression (\ref{43}) consider, for simplicity, the
contribution of the canonical pair $(\hat\rho^\tau, \hat
P^{(\rho)}_\tau)$ to the anticommutator (\ref{43}) of the BRST operator
(\ref{42}) using its different representations by the left and right
hand sides of (\ref{42}). For our purpose it is sufficient to
consider the contribution to the square of the BRST operator
proportional to the product of ghost operators $\hat C^{(W)}$ and
$\hat C^M$. Then for the BRST operator $\hat{\bf\Omega}$ (\ref{39'}) in
the l.h.s. of (\ref{42}) we obtain
\begin{equation}\label{43'}
\begin{array}{c}
\frac14\{\hat C^{(W)}(\hat\rho^\tau\hat P^{(\rho)}_\tau+\hat P^{(\rho)}_\tau\hat\rho^\tau)(\vec\sigma),-\hat C^M(\hat\rho^\tau\partial_{M'}\hat P^{(\rho)}_\tau+\partial_{M'}\hat P^{(\rho)}_\tau\hat\rho^\tau)(\vec\sigma')+
\\[0.2cm]
\hat C^M(\partial_{M'}\hat C^{(W)}\hat{\breve P}{}^{(W)}-\hat{\breve P}{}^{(W)}\partial_{M'}\hat C^{(W)})(\vec\sigma')\}.
\end{array}
\end{equation}
The desired contributions come from the two anticommutators in (\ref{43'})
\begin{equation}\label{47}
\begin{array}{c}
\quad\frac14\{\hat C^{(W)}(\hat\rho^\tau\hat P^{(\rho)}_\tau+\hat P^{(\rho)}_\tau\hat\rho^\tau)(\vec\sigma),-\hat C^M(\hat\rho^\tau\partial_{M'}\hat P^{(\rho)}_\tau+\partial_{M'}\hat P^{(\rho)}_\tau\hat\rho^\tau)(\vec\sigma')\}\\[0.2cm]
=-\frac{i}{2}\hat C^{(W)}(\vec\sigma)\hat C^M(\vec\sigma')(\hat\rho^\tau(\vec\sigma')\hat P^{(\rho)}_\tau(\vec\sigma)+\hat P^{(\rho)}_\tau(\vec\sigma)\hat\rho^\tau(\vec\sigma'))\partial_{M'}\delta^p(\vec\sigma-\vec\sigma')
\\[0.2cm]
+\frac{i}{2}\hat C^{(W)}(\vec\sigma)\hat C^M(\vec\sigma')(\hat\rho^\tau(\vec\sigma)\partial_{M'}\hat P^{(\rho)}_\tau(\vec\sigma')+\partial_{M'}\hat P^{(\rho)}_\tau(\vec\sigma')\hat\rho^\tau(\vec\sigma))\delta^p(\vec\sigma-\vec\sigma').
\end{array}
\end{equation}
 Transforming the arguments of the multipliers in front of the partial derivative $ \partial_{M'}\delta^p(\vec\sigma-\vec\sigma')$  to be coincident and equal  $\vec\sigma'$, we find the equivalent representation for the  r.h.s of Eq.(\ref{47}) as follows 
\begin{equation}\label{der}
-\frac{i}{2}\hat C^{(W)}\hat C^M(\hat\rho^\tau\hat P^{(\rho)}_\tau+\hat P^{(\rho)}_\tau\hat\rho^\tau)(\vec\sigma')\partial_{M'}\delta^p(\vec\sigma-\vec\sigma')
- \frac{i}{2}\partial_M\hat C^{(W)}\hat C^M(\hat\rho^\tau\hat P^{(\rho)}_\tau+\hat P^{(\rho)}_\tau\hat\rho^\tau)\delta^p(\vec\sigma-\vec\sigma').
\end{equation}

Correspondingly, for the second term we obtain
\begin{equation}\label{43'''}
\begin{array}{c}
\quad\frac14\{\hat C^{(W)}(\hat\rho^\tau\hat P^{(\rho)}_\tau+\hat P^{(\rho)}_\tau\hat\rho^\tau)(\vec\sigma),\hat C^M(\partial_{M'}\hat C^{(W)}\hat{\breve P}{}^{(W)}-\hat{\breve P}{}^{(W)}\partial_{M'}\hat C^{(W)})(\vec\sigma')\}\\[0.2cm]
=\frac{i}{2}\partial_M\hat C^{(W)}\hat C^M(\hat\rho^\tau\hat P^{(\rho)}_\tau+\hat P^{(\rho)}_\tau\hat\rho^\tau)\delta^p(\vec\sigma-\vec\sigma').
\end{array}
\end{equation}
The sum of the contributions (\ref{der}) and (\ref{43'''}) is the total divergence
\begin{equation}
\frac{i}{2}\partial_M\left(\hat C^{(W)}\hat C^M(\hat\rho^\tau\hat P^{(\rho)}_\tau+\hat P^{(\rho)}_\tau\hat\rho^\tau)\right)\delta^p(\vec\sigma-\vec\sigma')
\end{equation}
after the transfer of the derivative $\partial_{M'}$ from the $\delta$-function to the phase-space operators that is a correct operation for distributions.

Thus for the Hermitian representation, the contribution of the canonical
pair $(\hat\rho^\tau,\hat P^{(\rho)}_\tau)$ entering the Weyl
generator to the anticommutator $\{\hat
C^{(W)}\hat\Delta^{ext}_W(\vec\sigma),\hat C^M\hat L^{ext}_M(\vec\sigma')\}$
vanishes in agreement with the nilpotency of the BRST operator.
The same result can be found for the contributions of other
canonical pairs.

Now we are  interested in the same contribution to the square of the BRST operator proportional to the product of $\hat C^{(W)}$ and $\hat C^M$ ghosts, but in the $QP$-ordered and singular realization given by 
\begin{equation}\label{51}
\{\left(\hat C^{(W)}\hat\rho^\tau\hat P^{(\rho)}_\tau-i\hat C^{(W)}\delta^p(0)\right)(\vec\sigma),-\left(\hat C^M\hat\rho^\tau\partial_{M'}\hat P^{(\rho)}_\tau
-\hat C^M\partial_{M'}\hat C^{(W)}\breve P^{(W)}\right)(\vec\sigma')\},
\end{equation}
where we omitted the singular terms proportional to $\hat C^M\partial_M\delta^p(0)$ in the r.h.s. of the anticommutator  (\ref{51}) which do not contribute 
into the anticommutator. Then for the r.h.s. of Eq. (\ref{42}) we obtain the 
three summands
\begin{equation}
\begin{array}{c}
\quad\{(\hat C^{(W)}\hat\rho^\tau\hat P^{(\rho)}_\tau)(\vec\sigma),-(\hat C^M\hat\rho^\tau\partial_{M'}\hat P^{(\rho)}_\tau)(\vec\sigma')\}
\\[0.2cm]
=-i(\hat C^{(W)}\hat C^M\hat\rho^\tau\hat P^{(\rho)}_\tau)(\vec\sigma')\partial_{M'}\delta^p(\vec\sigma-\vec\sigma')
-i\partial_M\hat C^{(W)}\hat C^M\hat\rho^\tau\hat P^{(\rho)}_\tau\delta^p(\vec\sigma-\vec\sigma'),\\[0.2cm]
\quad\{(\hat C^{(W)}\hat\rho^\tau\hat P^{(\rho)}_\tau)(\vec\sigma),(\hat C^M\partial_{M'}
\hat C^{(W)}\hat{\breve P}{}^{(W)})(\vec\sigma')\}
=i\partial_M\hat C^{(W)}\hat C^M\hat\rho^\tau\hat P^{(\rho)}_\tau\delta^p(\vec\sigma-\vec\sigma').
\end{array}
\end{equation}
The sum of the above two summands is again the total divergence
\begin{equation}
i\partial_M\left(\hat C^{(W)}\hat C^M\hat\rho^\tau\hat P^{(\rho)}_\tau\right)\delta^p(\vec\sigma-\vec\sigma'),
\end{equation}
which can be omitted for the discussed model of the closed $p$-brane. So, the value of the anticommutator is determined by the singular term $i\hat C^{(W)}\delta^p(0)$ contribution equal
\begin{equation}\label{54}
\quad\{-i\hat C^{(W)}(\vec\sigma)\delta^p(0),(\hat C^M\partial_{M'}\hat C^{(W)}\hat{\breve P}{}^{(W)})(\vec\sigma')\}=-\delta^p(0)\hat C^M\partial_M\hat C^{(W)}\delta^p(\vec\sigma-\vec\sigma').
\end{equation}
One can see that the term does not transform to the total divergence.

Analogous calculations of the contributions of other canonical pairs entering the Weyl generator  will also give the same divergent answer modulo numerical coefficients. Summarizing all that contributions we arrive at the first summand in the r.h.s. of (\ref{43}).

Similarly, calculation of the (anti)commutators of the $OSp(1|8)$ operators (\ref{41}) with $\hat{\bf\Omega}$ (\ref{42}) gives the relations
\begin{equation}\label{44}
\begin{array}{rl}
\{\hat{\bf\Omega},\hat{\widetilde{\mathcal S}}_{\gamma}\}=0=&2i\int d^p\sigma(\hat C_\gamma+\hat C^M\partial_M\hat\theta_\gamma)\delta^p(0),\\[0.2cm]
\{\hat{\bf\Omega},\hat{\widetilde{\mathcal S}}_{\dot\gamma}\}=0=&-2i\int d^p\sigma(\hat{\bar C}_{\dot\gamma}-\hat C^M\partial_M\hat\theta_{\dot\gamma})\delta^p(0),\\[0.2cm]
[\hat{\bf\Omega},\hat{\widetilde{\mathcal K}}_{\gamma\lambda}]=0=&-2i\int d^p\sigma[(\hat C_\gamma\hat\theta_\lambda+\hat\theta_\gamma\hat C_\lambda)
+\frac{i}{2}\hat C^{M}\partial_M\hat z_{\gamma\lambda}]\delta^p(0),
\\[0.2cm]
[\hat{\bf\Omega},\hat{\widetilde{\mathcal K}}_{\gamma\dot\gamma}]=0=&-2i\int d^p\sigma[(\hat C_\gamma\hat{\bar\theta}_{\dot\gamma}-\hat\theta_\gamma\hat{\bar C}_{\dot\gamma})+\frac{i}{2}\hat C^{M}\partial_M\hat x_{\gamma\dot\gamma}]\delta^p(0).
\end{array}
\end{equation}
The conditions (\ref{43}) and (\ref{44}) contain the divergent terms proportional to $\delta^p(0)$ and $\partial_M\delta^p(0)$, so to attach meaning to them one should assign values to the singular limits of $\delta^p(\vec\sigma-\vec\sigma')|{}_{\vec\sigma=\vec\sigma'}$ and its derivatives, i.e. to make their regularization. In accordance with the regularization the (anti)commutation relations (\ref{35}) have to be changed by the regularized relations
\begin{equation}\label{35'}
[{\hat{\cal P}}^{\mathfrak M}(\vec\sigma),{\hat{\cal Q}}_{\mathfrak
N}(\vec\sigma')\}=-i\delta^{\mathfrak M}_{\mathfrak
N}\delta^p(\vec\sigma,\vec\sigma'),
\end{equation}
where the distribution $\delta^p(\vec\sigma,\vec\sigma')$ is a regularized image of the Dirac delta function $\delta^p(\vec\sigma-\vec\sigma')$. The substitution of the distributions $\delta^p(\vec\sigma,\vec\sigma')$ and $\partial_M\delta^p(\vec\sigma,\vec\sigma')$ in eqs.  (\ref{43}) and (\ref{44}) results in their solutions
\begin{equation}\label{48}
\delta^p(\vec\sigma,\vec\sigma)=0,\quad\partial_M\delta^p(\vec\sigma,\vec\sigma)=0.
\end{equation}
The regularization that requires $\delta^p(\vec\sigma,\vec\sigma)=0$ is well known in the physical literature (see, for instance, \cite{Gitman}). This regularization prescription leads to the closure of the
$\hat{\mathcal Q}\hat{\mathcal P}$-ordered and regularized
superalgebra of local and global symmetries of the considered
brane model. The regularization (\ref{48}) removes the problem of
ordering in the operators obtained from the classical expressions
and differing from each other by the terms proportional to
$\delta^p(0)$ and/or $\partial_M\delta^p(0)$. It shows the
possibility of quantization of the super $p$-brane model in this
regularization.

\section{\bf Conclusion}

The general problem of quantum brane realization of the
BPS states preserving $\frac{{\sf M}-1}{{\sf M}}$ fraction of partially spontaneously broken global $N=1$ supersymmetry at the classical level was analyzed on example of the twistor-like p-brane in four-dimensional space-time.
Twistor-like brane models are characterized by a fine tuning of a large set of classical local and global symmetries caused by the absence of tension.
We constructed classical BRST charge and generators of these global and gauge symmetries and proved the closure of their unified P.B. superalgebra.
 The P.B. realization of the nilpotency condition for the BRST charge and its (anti)commutativity with the unified symmetry generators were proved.
After that we considered quantization of the model and proved the preservation of the above classical results using  Hermitian operator realization of the symmetry generators and BRST operator.
This Hermitian realization gives the relevant physical foundation for the solution of the quantization problem. The remaining problem here is the construction of the vacuum and the Hilbert space of quantum states of tensionless super $p$-branes.

Otherwise, such a type problem was earlier considered in \cite{GLSSU}
using the general construction of the finite inner  product
\cite{HM} and the $\hat{\mathcal Q}\hat{\mathcal P}$-ordering in
the quantum BRST charge and generators of the symmetry algebra of
bosonic tensionless string. There was shown the existence of the
full physical vacuum that is annihilated by the  BRST and string
momentum operators. The latter condition  has picked up the
$\hat{\mathcal Q}\hat{\mathcal P}$-ordered representation as a
physical one for  tensionless objects. Motivated by these results
we studied the $\hat{\mathcal Q}\hat{\mathcal P}$-ordered
operator  realization of the brane symmetry generators and the
BRST operator. It was found
 the closure of the quantum generalized superconformal algebra
$OSp(1|8)$ in this realization and the (anti)commutativity of the quantum $OSp(1|8)$
generators with nilpotent BRST operator, because of the linearity
of the quantum operators in the momentum operators $\hat{\cal
P}^{\mathfrak M}$.  But, the $\hat{\mathcal Q}\hat{\mathcal
P}$-ordered representation for the $OSp(1|8)$ generators $\hat{\widetilde G}_{qp}$ and the
BRST operator $\hat\Omega$ turn out to be nonHermitian. We remind that the latter problem might be solved by the definite choice of the regularization assumption for the delta function $\delta^p(0)$ and its derivatives $\partial_M\delta^p(0)$. It shows a possibility to quantize the classical super $p$-brane preserving $\frac34$ fraction of partially spontaneously broken $D=4$ $N=1$ supersymmetry in the special regularization.

Our consideration admits straightforward generalization to the
case of the space-time dimensions $D$ belonging to the set
$2,3,4(mod8)$ consistent with the Majorana spinor existence.
Taking into account that the number coefficients in front of the
divergent terms containing $\delta^p(0)$ and
$\partial_M\delta^p(0)$ depend on $D$ one can hope that they might
 be equal to zero for some $D$ from the above mentioned set.
It would prove the existence of the quantized model outside the
regularization prescription. The above mentioned problems are
under our investigation.

\section{\bf Acknowledgements}

The authors are grateful to Jerzy Lukierski for inviting this contribution and for the warm hospitality at the University of Wroclaw.  
A.Z. thanks Fysikum at the Stockholm University and the Mittag-Leffler Institute for kind hospitality and I. Bengtsson, V. Gershun, S. Hwang, M. Kontsevich, O. Laudal, U. Lindstr\" om, M. Movshev, D. Piontkovskii, A. Rashkovskii and  B. Sundborg for useful discussions.  The work was partially supported by the grant of the Royal Swedish Academy of Sciences and by the SFFR of Ukraine under Project 02.07/276.

\end{document}